\documentclass[preprint2]{aastex}
\usepackage{amssymb,multirow,color}

\usepackage{amssymb,amsmath,amsfonts,graphicx,color,epsf}

\begin{document}

\title{Distribution of Magnetic Discontinuities in the Solar Wind and in MHD Turbulence}


\author{Vladimir Zhdankin$^1$, Stanislav Boldyrev$^1$, and Joanne Mason$^2$} 
\affil{$^1$ Department of Physics, University of Wisconsin-Madison \\
1150 University Avenue, Madison, Wisconsin 53706, USA}
\affil{$^2$ Department of Astronomy and Astrophysics, University of Chicago \\
5640 South Ellis Avenue, Chicago, Illinois 60637, USA}



\begin{abstract}
The statistical properties of magnetic discontinuities in the solar wind are investigated by measuring fluctuations in the magnetic field direction, given by the rotation $\Delta\theta$ that the magnetic field vector undergoes during time interval $\Delta{t}$. We show that the probability density function for rotations, $P(\Delta\theta)$, can be described by a simple model in which the magnetic field vector rotates with a relative increment $\Delta{B}/B$ that is lognormally distributed. We find that the probability density function of increments, $P(\Delta{B}/B)$, has a remarkable scaling property: the normalized variable $x=(\Delta{B}/B)\cdot(\Delta{t}/\Delta{t}_0)^{-\alpha}$ has a universal lognormal distribution for all time intervals $\Delta{t}$. We then compare measurements from the solar wind with those from direct numerical simulations of magnetohydrodynamic (MHD) turbulence. We find good agreement for $P(\Delta\theta)$ obtained in the two cases when the magnetic guide-field to fluctuations ratio $B_0/b_{rms}$ is chosen accordingly. However, the scale invariance of $P(\Delta{B}/B)$ is broken in the MHD simulations with relatively limited intertial interval, which causes $P(\Delta\theta)$ to scale with measurement interval differently than in the solar wind.
\end{abstract}


\section{Introduction} 

Since the beginning of the Space Age, the solar wind has been used as a natural laboratory for studying plasma turbulence. It is now widely accepted that the solar wind contains coherent magnetic structures \citep{veltri1999,sorriso-valvo1999, bruno2007, li2011, miao2011}. Observations have shown that magnetically dominated structures are ubiquitous in the heliosphere and are advected by both the fast and slow wind \citep{bruno2007b}. There have been many approaches to studying these structures \citep{li2008, greco2010, servidio2011}, hinging on the fact that they are typically associated with rapid spatial variation or reversal of the local magnetic field. In other words, structures are associated with discontinuities in the magnetic field. 

Although it is accepted that magnetic structures exist in the solar wind, there remains a debate regarding the origin and nature of these structures. Two primary theories have been put forth. One model envisions that the structures are flux tubes that originate in the solar corona and are passively advected by the solar wind. A second model contends that magnetohydrodynamic (MHD) turbulence dynamically produces intermittent structures as the solar wind flows outward. Although these two models are not mutually exclusive, it is important to determine to what degree each mechanism contributes.

An critical step toward answering this question was taken by \cite{borovsky2008}. He noted that magnetic discontinuities are implied by large fluctuations in the magnetic field direction across short measurement scales \citep{bruno2001,bruno2004}, which provides a convenient and statistically robust way to quantify magnetic discontinuities. Borovsky therefore considered fluctuations in the magnetic field direction, given by rotations $\Delta\theta = \cos^{-1}{( \boldsymbol{B}_1 \cdot \boldsymbol{B}_2 / | \boldsymbol{B}_1 | | \boldsymbol{B}_2 | )}$ where $\boldsymbol{B}_1$ and $\boldsymbol{B}_2$ are the magnetic field vectors for two measurements separated by time $\Delta{t}$. Using measurements from the ACE spacecraft, he studied the statistical properties of rotations by constructing a probability density function (pdf), $P(\Delta\theta)$. Borovsky discerned two populations in the pdf. The first population consists of strong discontinuities at $30^\circ < \Delta \theta < 170^\circ$ with an exponentially decaying pdf proportional to $\exp(-\Delta\theta/24.4^\circ )$. The second population consists of weak fluctuations at $5^\circ < \Delta \theta < 30^\circ$ which can be fit by exp($-\Delta\theta/9.4^\circ$). Subsequent anaylsis by \cite{miao2011} using data from the Ulysses spacecraft was in broad agreement with Borovsky's result. The interpretation put forth by these authors was that the first population represents strong discontinuities across the walls of coronal flux tubes, while the second population represents turbulent fluctuations. 

However, it is known that MHD turbulence can spontaneously produce strong magnetic discontinuities \citep{goldstein1995,boldyrev2006,greco2008}. These discontinuities correspond to intermittent structures, which primarily take the form of current sheets for inertial range turbulence. It has also been demonstrated that many statistical properties of the solar wind fluctuations are consistent with MHD turbulence. These include, for instance, the energy spectra of magnetic and velocity field fluctuations \citep{podesta2007,boldyrev2011}, the anisotropic scaling of structure functions \citep{chen2011}, and the waiting-time analysis of field increments \citep{greco2009,greco2010}. The MHD description is known to break down at small scales, but the solar wind fluctuations are then thought to be consistent with kinetic models of turbulence \citep{medvedev1997,howes2011}. We have also shown in a separate paper that MHD turbulence is able to reproduce the observed exponential tail of $P(\Delta\theta)$ for reasonable strengths of the magnetic guide-field \citep{zhdankin2012}. These facts support the picture where the population of discontinuities is governed by MHD turbulence, which may be complementary or alternative to interpretations by \citet{borovsky2008,miao2011}.

In this Letter, we investigate the statistical properties of magnetic discontinuities in the solar wind. We show that $P(\Delta\theta)$ is a consequence of the magnetic field vector having a strong tendency to undergo {\em pure rotations} with lognormally distributed relative increments $\Delta{B}/B$. This picture is attractive because the magnetic field in the solar wind is known to exhibit lognormal statistics to a good approximation \citep{burlaga2001}. Furthermore, we find that the pdf's of relative increments, $P(\Delta{B}/B)$, for different separations $\Delta{t}$ can be rescaled to match, that is, the distribution of $(\Delta B/B)\cdot(\Delta t/\Delta{t}_0)^{-\alpha}$ with appropriately chosen $\Delta t_0$ and $\alpha$ is a universal lognormal distribution. This strong scaling property implies the so-called monofractality, or absense of intermittency in the fluctuations of $\Delta B/B$. 

We then directly compare the statistical properties of rotations in the solar wind with numerical simulations of MHD turbulence. We find a good agreement in $P(\Delta\theta)$ for the two cases for reasonable ratios of simulation guide-field to fluctuations. We also find that the scaling of the distribution function of $\Delta{B}/B$ is not well preserved in the simulations and that the magnetic field vector does not have a strong tendency to undergo pure rotations. These differences may be the consequence of a much smaller inertial range in numerical simulations than in the solar wind. This causes the pdfs in the two cases to scale differently with measurement interval, and hence separations in the two cases cannot be directly compared.

\section{Analysis of solar wind}

First, we consider the properties of magnetic discontinuities in the solar wind. We analyze a time-series of spacecraft magnetic field measurements, $\boldsymbol{B}(t)$, taken by the Wind spacecraft over the years 2005-2010. Each vector is averaged over a time of $\Delta{t}_{min} = 3$ seconds, but the results are independent of the averaging scale as long as $\Delta{t} > \Delta{t}_{min}$. The analysis does not exclude events such as CMEs and CIRs. It also does not differentiate between slow and fast winds, which are known to have different typical properties and physical processes. A similar statistical study applied separately to the regions of fast or slow wind may in principle produce different results. 

To quantify magnetic discontinuities, we consider fluctuations in the magnetic field direction, given by the rotation $\Delta\theta$. The rotation $\Delta\theta$ between two magnetic field vectors, $\boldsymbol{B}_1$ and $\boldsymbol{B}_2$, separated by time interval $\Delta{t}$, is
\begin{align}
\Delta\theta = \cos^{-1}{ \left({\boldsymbol{b}_1 \cdot \boldsymbol{b}_2} \right) }, \label{eq:angle}
\end{align}
where ${\boldsymbol b}={\boldsymbol B}/B$ is a unit vector along the magnetic field. A large rotation implies a magnetic discontinuity, while a small rotation represents weak fluctuations. The statistical properties of rotations then provide a window into understanding the nature of discontinuities and intermittency. Similar measurements of magnetic field rotations have been applied by \cite{perri2009} to study anisotropy.

\begin{figure}
\plotone{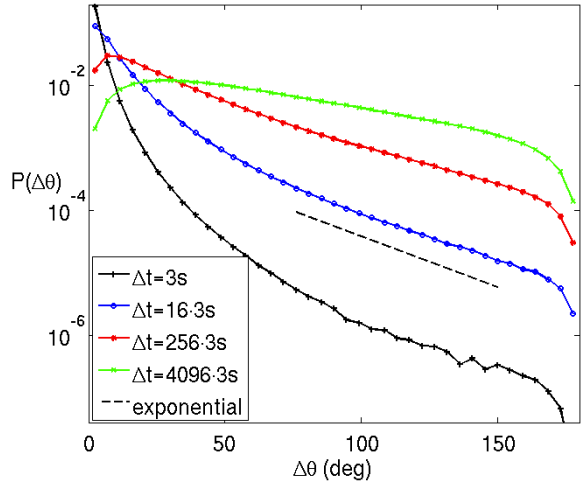}
\caption{The pdf of rotations in the magnetic field direction, $P(\Delta\theta)$, for the solar wind with measurement intervals of $\Delta{t} \in \{1,16,256,4096\}\cdot{3}$ seconds (measured by the WIND spacecraft). While part of the tail ($\Delta\theta > 60^\circ$) may be fit by an exponential function (straight line in the plot), the bulk of the pdf ($\Delta\theta < 60^\circ$) varies drastically with $\Delta{t}$. \label{fig:wind}}
\end{figure}

In Fig.~\ref{fig:wind}, the pdf of rotations, $P(\Delta\theta)$, is shown for $\Delta{t} \in \{1,16,256,4096\}\cdot{3}$ seconds. The pdf has a strong dependence on $\Delta t$. The tail ($\Delta\theta > 60^\circ$) can be approximately fit by an exponential function, while the bulk of the pdf ($\Delta\theta < 60^\circ$) varies drastically. This reveals that the resolution used to measure fluctuations is an important parameter. This is reasonable because the correlation between measurements is increased when $\Delta{t}$ is decreased, which in turn increases the likelihood of small $\Delta\theta$. 

\begin{figure}
\plotone{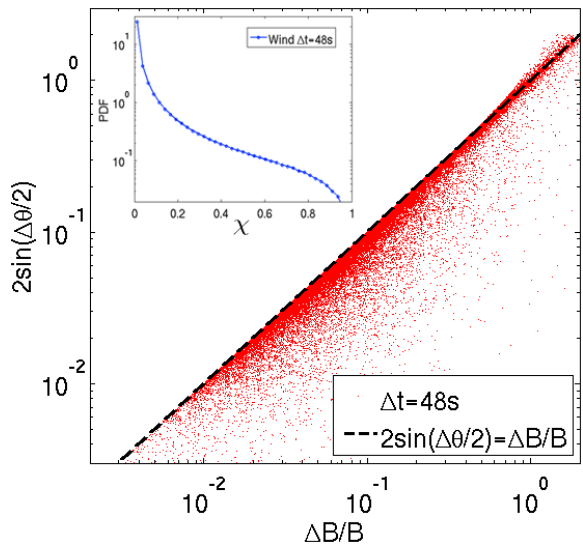}
\caption{A scatterplot of $2\sin{(\Delta\theta/2)}$ versus $\Delta{B}/{B}$, which are equal if the magnetic field vector undergoes a pure rotation between measurements. Also shown, in the inset, is the pdf of the normalized difference $\chi = |\Delta{B}/B - 2\sin{(\Delta\theta/2)}|/(\Delta{B}/B)$, which is strongly peaked at zero. These plots show that the magnetic field in the solar wind tends to rotate between the measurements rather than change in strength. \label{fig:rotation}}
\end{figure}

We now show that $P(\Delta\theta)$ can be described by a simple model. 
We define the relative increment in magnetic field as $\Delta{B}/B = |\boldsymbol{B}_2 - \boldsymbol{B}_1|/|\boldsymbol{B}_1|$. If the magnetic field vector undergoes a pure rotation between the measurements, that is, does not change amplitude, then $\Delta B/B$ depends only on the rotation angle $\Delta\theta$, satisfying
\begin{align}
{\Delta{B}}/{B} = \vert {\boldsymbol b}_2 -{\boldsymbol b}_1\vert =2 \sin{(\Delta\theta/2)}, \label{eq:rotation}
\end{align}
where fluctuations must lie within $0 < \Delta{B}/B < 2$ to be meaningful. We find that the magnetic field vector in the solar wind indeed strongly satisfies Eq.~\ref{eq:rotation}, as shown in Fig.~\ref{fig:rotation}. Therefore, between two closely spaced measurements ($\Delta{t} \lesssim 10^4$ seconds), the magnetic field tends to mostly rotate instead of change in strength, and the fluctuations in the rotation angle can be studied in terms of the relative increments $\Delta B/B$.

\begin{figure}
\plotone{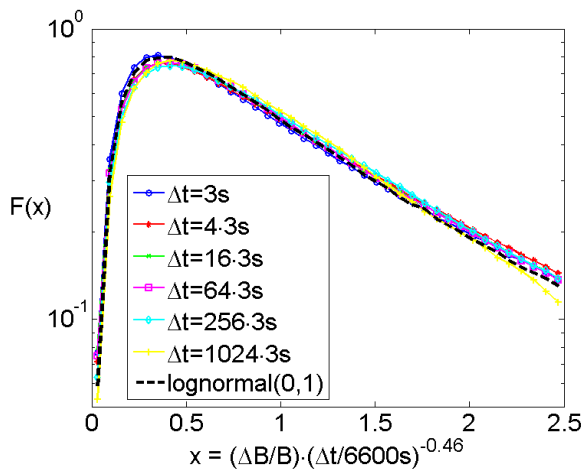}
\caption{The quantity $x = (\Delta{B}/B)\cdot(\Delta{t}/\Delta{t}_0)^{-\alpha}$ has the pdf $F(x)$ that is independent of $\Delta{t}$, where $\Delta{t}_0 \approx 6.6 \times 10^3$ seconds and $\alpha \approx 0.46$. The curves shown for $\Delta{t} \in \{1,4,16,64,256,1024\}\cdot 3$ seconds are all in remarkable agreement with the lognormal distribution of $\mu = 0$ and $\sigma = 1$, indicating that $\Delta{B}/B$ is scale-invariant over the given values of $x$. For small $\Delta{t}$, the agreement also extends far beyond the values of $x$ shown, almost up to the cutoff at $\Delta B/B=2$. \label{fig:rescale}}
\end{figure}

We find that $P(\Delta{B}/B)$ in the solar wind is well fit by a lognormal distribution for $0 < \Delta{B}/B < 2$. The lognormal distribution is given by 
\begin{align}
f(x) &= \frac{1}{x \sigma \sqrt{2\pi}} \exp{\left(-\frac{1}{2\sigma^2}(\log{x} - \mu)^2 \right)},
\end{align}
where $\mu$ and $\sigma$ are the location parameter and scale parameter, respectively. The fit is not surprising since other measurements of magnetic fields in the solar wind also have approximately lognormal statistics, e.g. \cite{burlaga2001}. The lognormal fits to $P(\Delta{B}/B)$ can all be chosen with $\sigma = 1$, which implies that $P(\Delta{B}/B)$ for any $\Delta{t}$ can be rescaled into the universal lognormal function according to the formula
\begin{eqnarray}
P\left(\frac{\Delta B}{B}; \Delta t\right)= \left(\frac{\Delta t}{\Delta t_0}\right)^{-\alpha}F\left(\frac{\Delta B}{B}\left(\frac{\Delta t}{\Delta t_0}\right)^{-\alpha}\right),
\label{scaling}
\end{eqnarray}
where $F(x)$ is the universal lognormal probability density function, 
\begin{align}
F(x) &= \frac{1}{x \sqrt{2\pi}} \exp{\left(-\frac{1}{2}\log^2{x} \right)},
\end{align}
and the best fit is given by $\Delta t_0\approx 6.6 \times 10^3$ seconds and $\alpha \approx 0.46$. This property is illustrated in Fig.~\ref{fig:rescale}, where the pdfs of the rescaled variable $x = (\Delta{B}/B)\cdot(\Delta{t}/\Delta{t}_0)^{-\alpha}$ are in remarkable agreement for $3 < \Delta{t} < 1024\cdot3$ seconds. We found that the measured curves follow the scale-invariant lognormal form almost until the cutoff that occurs at $\Delta{B}/B = 2$. For different values of $\Delta{t}$ such a cufoff corresponds to different values of $x$. Fig.~\ref{fig:rescale} shows the range of $x$, where all the measured curves follow the lognormal form~$F(x)$.

This scaling property of the probability density function implies that the field $\Delta B/B$ has a scale-invariant and nonintermittent (monofractal) distribution. In addition, it demonstrates that there exists a certain scale $\Delta{t}_0$, which can probably be identified with the outer scale of turbulence.

\begin{figure}
\plotone{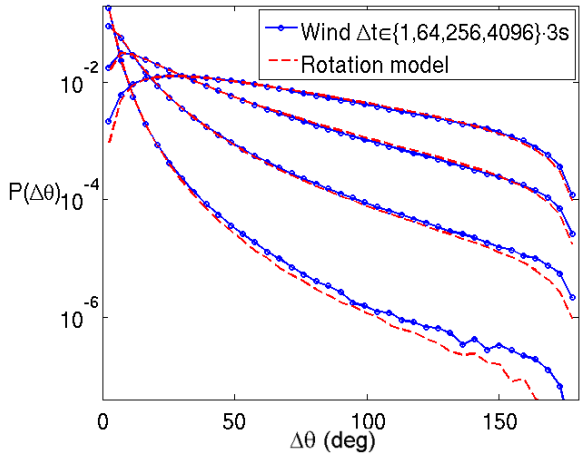}
\caption{Comparison of the pdf for rotations measured in the solar wind (in blue) and the pdf derived from lognormal $P(\Delta{B}/B)$ with the assumption that vectors undergo pure rotations (in red). The separations shown are $\Delta{t} \in \{1,16,256,4096\}\cdot3$ seconds. \label{fig:model}}
\end{figure}

We therefore model the rotations $\Delta\theta$ as follows. Assume that $P(\Delta{B}/B)$ is lognormal and scales with $\Delta{t}$ as described by (\ref{scaling}), and also assume that the vector undergoes a pure rotation so (\ref{eq:rotation}) is applicable. Then $P(\Delta\theta)$ is obtained. This model gives excellent agreement with the pdf of rotations in the solar wind, as shown in Fig.~\ref{fig:model} for $\Delta{t} \in \{1,16,256,4096\}\cdot3$ seconds. It is worth noting that the bulk of $P(\Delta\theta)$ is also approximately lognormal, in agreement with previous conclusions by \cite{vasquez2007} and \cite{bruno2004}. We have however shown that modeling $\Delta B/B$, rather than $\Delta \theta$, from a lognormal distribution better fits the data. The weak fluctuations and strong discontinuities characterized by $\Delta\theta$ may then be regarded as a single population described by lognormal fluctuations in the magnetic field and a tendency of the magnetic field vector to rotate.

\section{Analysis of MHD turbulence}

We now discuss the statistical properties of rotations in MHD turbulence, and compare them to the solar wind case. We use data from direct numerical simulations of 3D incompressible strong MHD turbulence. The MHD equations are
\begin{eqnarray}
\partial_t \boldsymbol{v} + (\boldsymbol{v} \cdot \nabla) \boldsymbol{v} &=& - \nabla p + (\nabla \times \boldsymbol{B}) \times \boldsymbol{B} + \nu \nabla^2 \boldsymbol{v} + {\bf f}_1, \nonumber \\
\partial_t \boldsymbol{B} &=& \nabla \times (\boldsymbol{v} \times \boldsymbol{B}) + \eta \nabla^2 \boldsymbol{B} +{\bf f}_2, \nonumber  \\
\nabla \cdot \boldsymbol{v} &=& 0, \nonumber \\
\nabla \cdot \boldsymbol{B} &=& 0,
\end{eqnarray}
where $\boldsymbol{v}(\boldsymbol{x},t)$ is the plasma velocity, $\boldsymbol{B}(\boldsymbol{x},t)={\bf B}_0+{\bf b}$ is the magnetic field that has both uniform (${\bf B}_0$) and fluctuating (${\bf b}$) components, $p$ is the pressure, and ${\bf f}(\boldsymbol{x},t)$ is the external forcing. We take the viscosity $\nu$ and resistivity $\eta$ to be equal.

The equations are solved on a triply periodic domain using standard pseudospectral methods. The time-advancement of the diffusive terms is carried out exactly using the integrating factor method, while the remaining terms are treated using a third-order Runge-Kutta scheme. For a detailed description of the numerical method, see, e.g.,~Ref.~\citep{cattaneo2003}. The turbulence is driven at the largest scales by applying random forces
${\bf f}_1$ and ${\bf f}_2$ in Fourier space at wave-numbers
$2\pi/L \leq k_{\perp} \leq 2 (2\pi/L)$, $k_\| =
2\pi/L$. The correlation between the forces is chosen as to mimic the driving by independent counter-propagating shear-Alfv\'en modes, see, e.g., \citep{perez2010,boldyrev2011}, however turbulence may also be driven by other choices of large-scale forcing; this does not affect the inertial interval, see~\citep{mason2008}. The forces have no component along $z$ and are solenoidal in the $xy$-plane.  All of the Fourier coefficients outside the above range of wave-numbers are zero and inside that range are Gaussian random numbers that are refreshed on average every $0.1L/(2\pi v_{rms})$ time units (that is, force is updated approximately 10 times per large-scale turnover time) with amplitudes chosen so that $v_{rms}\sim 1$. We conducted a number of MHD simulations with different ratios of $B_0/b_{rms}$. The analysis is performed on several statistically independent snapshots corresponding to a steady-state, with resolution of $1024^3$ and Reynolds number $Re \approx 3000$.

For a fixed time snapshot, we measure the rotation of the magnetic field between points separated by spatial distance $\Delta{x}$. Therefore $\boldsymbol{B}_1$ and $\boldsymbol{B}_2$ in Eq.~\ref{eq:angle} are taken from a magnetic field profile $\boldsymbol{B}(x,y,z)$. We consider separations taken in the plane perpendicular to the guide-field, although the results are similar for separations along the guide-field. Using spatial separations in MHD instead of time separations is a significant difference from the solar wind analysis. However, we expect temporal measurements in the solar wind to be equivalent to spatial ones because of the Taylor hypothesis, $\Delta{t} \approx \Delta{x}/V_{SW}$, where $V_{SW}$ is the (approximately constant) solar wind velocity. Therefore, we should observe $\Delta{t} \propto \Delta{x}$ if the solar wind observations are consistent with MHD turbulence.

We find that $P(\Delta\theta)$ has a strong dependence on the ratio of magnetic guide-field to root mean square (rms) fluctuations, $B_0/b_{rms}$. Specifically, the characteristic angle for the exponential tail decreases with increasing $B_0/b_{rms}$, i.e. the slope on the log-linear axis becomes steeper \citep{zhdankin2012}. We find that choosing $B_0/b_{rms} \approx 0.32$ gives a good agreement with the solar wind measurements considered here, which is reasonable since $B_0 \sim b_{rms}$ in the solar wind. Therefore, we use MHD data with $B_0/b_{rms} \approx 0.32$ in the remaining analysis. We also find that $P(\Delta\theta)$ does not change significantly in simulations of lower Reynolds number ($Re \approx 2200$) and lower resolution ($512^3$), indicating that even higher Reynolds numbers may need to be achieved for better agreement with the observational data. 

\begin{figure}
\plotone{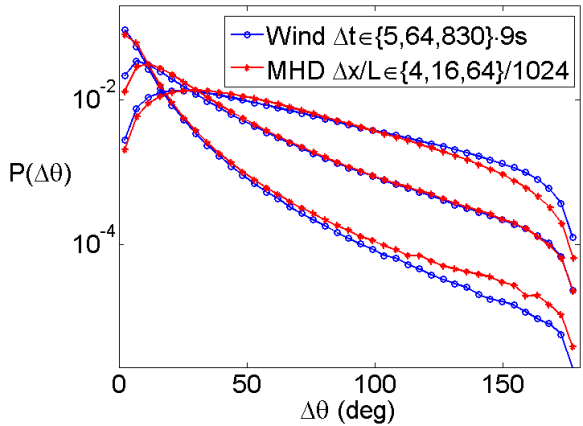}
\caption{A comparison of the pdfs of $\Delta\theta$ for MHD turbulence (in red) with $\Delta{x}/L \in \{4,16,64\}/1024$ and for the solar wind (in blue) with $\Delta{t} \in \{5,64,830\}\cdot{9}$ seconds, consistent with the relation $\Delta{t} \sim (\Delta{x})^{1.85}$. There is generally good agreement in the pdfs for both cases. \label{fig1}}
\end{figure}

A direct comparison of $P(\Delta\theta)$ for MHD turbulence and for the solar wind is shown in Fig.~\ref{fig1} for several measurement intervals. To get the best agreement between the two cases, we have chosen intervals of $\Delta{x}/L \in \{4,16,64\}/1024$ for MHD and $\Delta{t} \in \{5,64,830\}\cdot{9}$ seconds for the solar wind. This requires approximately that $\Delta{t} \propto \Delta{x}^{1.85}$ in order to get the solar wind pdfs to match the MHD pdfs, which seemingly violates the Taylor hypothesis. We find also that the scale invariance of $P(\Delta{B}/B)$ as described by (\ref{scaling}) is broken in our MHD simulations: although $P(\Delta{B}/B)$ is well fit by a lognormal distribution for all $\Delta{t}$, the fits require different values of $\sigma$ and therefore cannot be rescaled into one another. We also find that the magnetic fluctuations in the simulations are not dominated by rotations. 

These differences are likely caused by a much shorter inertial range (less than one decade) that spoils the scale invariance of $P(\Delta{B}/B)$ as described by (\ref{scaling}), which in turn modifies the scaling of $P(\Delta\theta)$. One, in princple, should not expect to see such a strong scaling property of $\Delta{B}/B$ in a rather modest simulation. A related difference is that the MHD measurements are near the dissipation range ($\Delta{x} \lesssim 16$), while the solar wind measurements are well within the intertial range. Therefore the simulations must have a larger separation of scales to more accurately capture the physics that describes the scaling. Nevertheless, we see that for any fixed separation, MHD is able to produce similar pdfs as in the solar wind, which shows that discontinuities in the solar wind are accurately reproduced by simulations. 

\section{Conclusions}

We found that the pdf of rotations in the magnetic field of the solar wind has a shape that is in good qualitative agreement with strong MHD turbulence (Fig.~\ref{fig1}). The numerical simulations are consistent with the observations if the magnetic guide-field to rms fluctuations ratio is $B_0/b_{rms}\approx 0.32$, which is reasonable for the solar wind. The approximately exponential tail of $P(\Delta\theta)$ is consistent with solar wind analysis done by \cite{borovsky2008} and \cite{miao2011}, but our interpretation of the result is different. Our results suggest that the pdf is closely associated with lognormal fluctuations of the magnetic field. Since such fluctuations are present in MHD simulations, this suggests that MHD turbulence can, to a large extent, describe the magnetic discontinuities observed in the solar wind. This supports a picture in which the discontinuities are associated with intermittent structures that arise from the MHD energy cascade \citep{sorriso-valvo2007,macbride2008}. This can be alternative or complementary to the coronal flux tube model of discontinuities. 

What remains to be better understood is the difference in scaling of $\Delta{x}$ and $\Delta{t}$ required to obtain agreement between the pdfs for the two cases. This result is surprising, and we cannot provide an exhaustive explanation with our current level of understanding and numerical simulations. Our results may suggest that the different scaling in MHD simulations arises due to a very limited inertial range relative to the solar wind. The fact that $P(\Delta\theta)$ are in such good agreement despite these differences suggests that there could be an underlying universal description for them.

\acknowledgements
This work was supported   
by the US DoE grants DE-FG02-07ER54932, DE-SC0003888, DE-SC0001794, the NSF Grant PHY-0903872, the NSF/DOE Grant AGS-1003451, and the NSF Center for Magnetic Self-organization in Laboratory and Astrophysical Plasmas at the University of  Wisconsin-Madison and the University of Chicago, as well as by an allocation of advanced computing resources at the National Institute for Computational Sciences.


\clearpage

\end{document}